# Trust Concerns in Health Apps collecting Personally Identifiable Information during COVID-19-like Zoonosis

Molla Rashied Hussein
*Dept. of Comp. Sci. and Engg.*
*University of Asia Pacific*
Dhaka, Bangladesh
mrh.cse@uap-bd.edu

Md. Ashikur Rahman
*Dept. of Comp. Sci. and Engg.*
*University of Asia Pacific*
Dhaka, Bangladesh
16201026@uap-bd.edu

Md. Jahidul Hassan Mojumder
*Dept. of Comp. Sci. and Engg.*
*University of Asia Pacific*
Dhaka, Bangladesh
16201035@uap-bd.edu

Shakib Ahmed
*Dept. of Comp. Sci. and Engg.*
*University of Asia Pacific*
Dhaka, Bangladesh
16201039@uap-bd.edu

Samia Naz Isha
*Dept. of Pub. Health & Primary Care*
*The University of Cambridge*
Cambridge, United Kingdom.
sni25@medschl.cam.au.uk

Shaila Akter
*Department of Global Health*
*Heidelberg University Hospital*
Heidelberg, Germany.
dr.shailaakter@gmail.com

Abdullah Bin Shams
*Dept. of Electrical and Comp. Engg.*
*University of Toronto*
Toronto, Canada
abdullahbinshams@gmail.com

Ehsanul Hoque Apu
*Dept. of Biomedical Engineering*
*Michigan State University*
East Lansing, MI, USA
hoqueapu@msu.edu

*Abstract*—Coronavirus disease 2019, or COVID-19 in short, is a zoonosis, i.e., a disease that spreads from animals to humans. Due to its highly epizootic nature, it has compelled the public health experts to deploy smartphone applications to trace its rapid transmission pattern along with the infected persons as well by utilizing the persons' personally identifiable information. However, these information may summon several undesirable provocations towards the technical experts in terms of privacy and cyber security, particularly the trust concerns. If not resolved by now, the circumstances will affect the mass level population through inadequate usage of the health applications in the smartphones and thus liberate the forgery of a catastrophe for another COVID-19-like zoonosis to come. Therefore, an extensive study was required to address this severe issue. This paper has fulfilled the study mentioned above needed by not only discussing the recently designed and developed health applications all over the regions around the world but also investigating their usefulness and limitations. The trust defiance is identified as well as scrutinized from the viewpoint of an end-user. Several recommendations are suggested in the later part of this paper to leverage awareness among the ordinary individuals.

## I. INTRODUCTION

The zoonotic novel disease named coronavirus disease 2019 (COVID-19) has changed the course of the global economy by putting constraints on individuals in daily commercial activities [1]. The highly contagious disease has potentially affected health facilities' capacity, even in developed countries where there are proven and robust healthcare systems [2]. To control and track the COVID-19 trajectory, different administrations have utilized modern artificial intelligence (AI) based methods integrated with 5G technology and aerial drone devices for real-time COVID-19 tracking [3][4]. With no specific vaccine or medicine for preventing or curing [5], currently, the best approach to avoid COVID-19 is not to get ourselves exposed to it [6]. Social distancing must be preserved between all, as recent studies [7] have shown that even an asymptomatic person may silently propagate the disease without raising any suspicion.

This social distancing induces the public health specialists and workers to collaborate with the technical researchers to deploy the health applications (apps) to monitor people digitally, as the manual process is quite tedious to execute. Instead of gathering the personally identifiable information (PII) of the affected individual along with contact information for the last 2-3 weeks through extensive interviews, the health apps in smartphones can be an excellent alternative that have already been introduced, developed, and even deployed in some countries [8].

Also, a majority of the people can use those apps using their smartphones not only in developed countries but also in the low- and middle-income countries (LMICs) [9]; proper app usability has to be guaranteed by comparing in terms of development as well as confidentiality along with popularity among the end-users, and last but not least, user-friendly interfaces.

However, as the PII is stored in those apps, potential exposure of the PII may cause a violation of the user's privacy. Moreover, the re-identification of a person can be done by only a few demographic data from the PII. Thus an individual may be targeted for a prospective cyber-attack or threat.

This paper is organized as follows: section I introduces the concept; the terminologies are defined in section II, the relevant health apps and their functionalities are reviewed in section III. After that, the trust concerns are summarized in section IV and various recommendations are suggested in section V. The paper concludes with section VI advising future implementation as per the recommendations.

## II. DEFINITION OF TERMINOLOGIES

Before discussing the health-related apps and their features, fundamental trust issues concerning cyber security and privacy must be briefly introduced. The first one is the concept of the semi-honest model, the second one is the activity of a malicious actor, and the last one is the possibility of re-identification of a person to track and carry on cyber-attack or threat.

In the real world, when a user is using smartphone apps during any type of epidemic situation for public health activities, some PII is saved into the storage of that application. Those app data can indicate where the users are going or whom the users are dealing with at any moment or even the current location of those users.

Now, after those epidemic situations go away, still the PII remains saved in that particular storage which the user does not want to. For that, the user demands the health app that keeps the data for a specific timeframe, and after that, those data will be removed. For that, users need a fully trustable model or a full honest model where there will be no hackers or malicious actors who are going to misuse the databases of a user. Still, it is impossible to create a full honest model because we cannot assume that all the actors in a security model behave in a completely honest manner [10].

However, that necessarily does not imply that a vast number of malicious actors are always active to harm the unsuspecting users all the time. Because even though they may wish to harm or cheat, they in fact, act rather in a semi-honest way [10]. Therefore, we are proposing for a semi honest model to get rid of this problem of having trust.

Semi honest party means the person or user who accepts all the terms and conditions of an application with the exception that it saves a list of all its intermediate computations. A semi-honest party always agrees with the "honest verifier" in the statement of empty-knowledge. To use smartphone apps for public health purposes, a user must register in it and share PII. In contrast, general malicious behavior may be non-changeable for many users, semi-honest behavior may be useful for them. Using a semi-honest model, users can be more reliable to use those contact tracing applications.

In a malicious model, there is not an ideal condition stable here. The malicious model means where all participants are in the attacking mode. The first thing that must follow is that there should be no force to enter the protocol. There are many kinds of malicious behaviors. A possible malicious behavior can be the failure to start the execution. Or it can break the commission at any desired point in time. One of the targets of this model is, in a one on one party communication, not to give a third party the access to it. It is known to us that full fairness is not achievable. Another possible malicious behavior shows that when considering malicious rivals, it is unknown what their input to the protocol is. It means a malicious party can enter the protocol with a piece of open information, which may not equal to its "true" local input. There is no possibility of getting to know the actual input. From those discussions, we can see that it is not possible at all to stop the malicious actors.

Behind a privacy model, there may be a significant number of threat actors. A threat actor means essence or person who has the intent to create an impact on the security and safety of others. In the cyber security and threat intelligence in a privacy model, the threat actors mean a massive number of any essences or persons either intentionally or unintentionally, who try to, and many times successfully execute malicious deeds against any industries.

Many kinds of Cyber Threat Actors Have Different types of Motivations. We can specify those thread actors mainly in two types which are: malicious actors and nefarious actors. Malicious actors are the ones who try to hamper the system whenever they get the chance. Nefarious actors, on the other hand, are the ones who are always present online and try to break the systems. In real-life, nefarious actors present in a privacy model are quite rare. Some malicious actors can still be present in a privacy model. Some examples of threat actors in a privacy model are given:

Cyber-terrorist: Actors who attack via technology in cyber-space are called cyber-terrorist. For a bad political intention, outsider groups of actors are using cyber techniques to influence the public for an evil motive and try to change the political state. They try to create fear in the public's mind. There are no official reports filed against them because of their unknown identity. So that after doing this type of crime, they started to do more such as learn how to build a bomb, to recruit, have a meeting, and connect with like-minded individuals to spread their propaganda.

Hacktivists: Hacking is not used for lousy intention all the time. Some organizations may hire several hackers to access any legal information for their business purpose. But sometimes a hacktivist is trying to embarrass celebrities politically or economically, to highlight human rights, waking up an organization to its sensitive things, to going after groups whose ideologies they disagree with. Hacktivists may steal and disseminate sensitive, proprietary, or, sometimes, classified data in the name of free speech. Other times, they aim to deny access to a particular service or website by conducting a distributed denial-of-service attack, essentially denying legitimate access by flooding a website with more traffic than it can handle, causing the site to crash.

State-sponsored actors: receive direction, funding, or technical assistance from a nation-state to advance that nation's interests. State-sponsored actors have stolen and infiltrated intellectual property, sensitive PII, and money to fund or further espionage and exploitation causes. In rare cases, these data appear for sale on underground black markets. Instead, these data are usually kept by the actors for their purposes. Although the data taken from data breaches might not always appear on underground markets, what can occur are the tools and guides for how to take advantage of the vulnerabilities that allowed access to the vulnerable systems in the first place. As an example, a researcher published the flaw that was used to penetrate Equifax, and within 24 hours, the information was posted to hacking websites and included in hacking toolkits.

Cybercriminals: Cybercriminals are mainly profit-driven and represent a lengthy period and worldwide threat. The target of the Cybercriminals is to sell the databases, hold for ransom, and otherwise absorb for financial gain. Cybercriminals may work individually or in groups to achieve their purposes [11].

Tracking: It generally means following the path or current location of delivery in real-time. Tracking is about gaining insights in real-time. A tracking app can, for example, determine a person's current location using geo-data (e.g., via GPS coordinates or radio cell location). If it additionally tracks who has been where and when it even allows creating detailed movement profiles.

Tracing: On the other side of the coin, it means following the path of a delivery backward from its current point to where it started. Tracing is all about gaining insights in retrospect. A tracing app can be used, for example, to trace physical contacts between people. Using Bluetooth, a technology that enables digital devices to communicate with each other over short distances, it can measure the distance between smartphones based on the strength of the radio signals and thus detect encounters between users (proximity tracing).

Demographic variables: They are defined as the independent variables as they cannot be manipulated. In research, demographic variables may be either categorical (e.g., gender, race, marital status, psychiatric diagnosis) or continuous (e.g., age, years of education, income, family size) [12].

De-identification:- The process of anonymizing datasets before sharing data. It has been the leading paradigm used in research and elsewhere to share data while preserving people's privacy.

In today's world, there is some modern data protection law, such as the European General Data Protection Regulation (GDPR) and the California Consumer Privacy Act (CCPA) consider that every person in a dataset has to be protected for the dataset to be considered anonymous. Data protection laws worldwide regarded as anonymous data as not personal data anymore.

III. RELATED WORKS

To secure their people from COVID-19, governments around the world have authorized the usage of a smartphone app designed and implemented by the local technical enterprises with/without the collaboration with the global tech giants. In this section, we will go by the regions, i.e., continents around the world.

Starting alphabetically from Africa, the continent has seen its country Ghana implementing the GH COVID-19 Tracker by crowd sourcing data to provide knowledge on the potential place of verified individuals to support health authorities to determine high-risk individuals. Crowd-sourced data collection is a sharing method of making a data-set with the support of a massive group of the crowd and using a more secured semi-honest model to improve Crowd-sourcing.

Next from Asia, more specifically, South Asia, India has developed an application named Aarogya Setu [13]. This application has got very good documentation [14]. Although there was no mention of disposal of data after the pandemic in the initial development, the updated privacy policy [15] states that all PII collected will be retained on the mobile device for 30 days from the date of the collection, after which, if it has not already been uploaded to the server, will be purged from the App. All PII uploaded to the server will, to the extent that such information relates to people who have not tested positive for COVID-19, will be purged from the server 45 days after being uploaded. All PII collected from persons who have tested positive for COVID-19 will be purged from the server 60 days after such persons have been declared cured of COVID-19.

The following Table 1 is going to summarize the merits and demerits of several health apps spread over the numerous countries under the regions/continents.

TABLE I. TRUST IN HEALTH APPS: MERITS AND DEMERITS

| Region | App features by region, country and name | | | |
|---|---|---|---|---|
| | *Country Name* | *App Name* | *Merits* | *Demerits* |
| Africa | Ghana | GH COVID-19 Tracker | Crowd-sourcing | Security Model |
| Asia | India | Aarogya Setu | Well documented privacy policy | Central server allows potential hacking of PII |
| | Malaysia | My-Sejahtera | DP3T | Data Policy |
| | Singapore | TraceTogether | Secure Server | Users can delete the app anytime |
| | Bahrain | BeAware | (not enough information) | Documentation |
| | Tetamman | Tetamman | (not enough information) | Power leakage |
| | Israel | Hamagen | Crosscheck GPS data with MOH data | Documentation |
| Europe | Czech Republic | e-Rouška | Self-explanatory | Linguistic barrier |
| | Hungary | VírusRadar | High security | (not enough information) |
| | Iceland | Rakning C-19 | Supervised by Health Directorate | (not enough information) |
| | Denmark | Smittestopp | Helpful FAQ information | Battery drainage |
| South America | Colombia | CoronApp | Well documented | No English version |
| Oceania | Australia | COVIDsafe | Well documented | (not enough information) |
| | New Zealand | Covid Tracer | Maximum security unless user misuse | User dependent action |

Moreover, Aarogya Setu [13] app alerts a user when they are nearby an infected person by using both Bluetooth and GPS. Several Data Science concepts, such as Classification, Association Rule Mining, and Clustering to analyze COVID-19 spread is being used [14]. As PII of numerous users of the app is stored on one server, this design potentially allows the potential malicious actors to hack the PII of users.

Meanwhile, from Asia, more specifically from Southeast Asia, MySejahtera [16] in Malaysia needs to mention good documentation describing its data privacy. Another app from Malaysia, MyTrace [17] accepts a community-driven approach where participating devices exchange proximity information whenever an app detects another nearby device with MyTrace installed. As this app is based on the DP3T (Decentralized Privacy-Preserving Proximity Tracing) algorithm [18], it does not require location permissions. As long as location permissions are required, this app serves as a state surveillance apparatus, not a tool to assist in breaking the chain.

On the other hand, Singapore has implemented TraceTogether with proper privacy statement [19]. The TraceTogether program enhances Singapore's contact tracing efforts in the fight against COVID-19. It comprises the TraceTogether app and the TraceTogether token. The app was released on 20 March, while the token was rolled out on 28 June. The timeline is as follows: 1 April 2020 - Clarified the collection of anonymized analytics data, 1 June 2020 - Clarified the collection of identification details and usage of data for contact tracing, 3 Sep 2020 - Included the TraceTogether Token in the Privacy Safeguards. Both the App and the Token can be used voluntarily, and it facilitates the contact tracing process.

With the user's consent, it exchanges encrypted and anonymized Bluetooth signals with nearby TraceTogether devices, and the Bluetooth data after 25 days are automatically deleted. This allows users to be informed if users were in prolonged physical proximity with an infected person. They are committed to safeguarding users' privacy and will only use users' data for contact tracing purposes. They store limited data. The only identity data they store is: users' contact/mobile number, users' identification details, a random anonymized User ID. When users sign up, a random User ID is generated and associated with users' contact/mobile number and identification details. Users' contact/mobile number, identification details, and User ID are stored in a secure server, and never shown to the public. They do not collect data about users' GPS location. TraceTogether uses Bluetooth to approximate users' distance to other TraceTogether devices. They do not collect data about users' GPS locations. Neither do they collect data about users' WiFi or mobile network. Data about devices near users do not reveal personal identities. When users are close to another TraceTogether device, both devices use Bluetooth to exchange a Temporary ID. This Temporary ID is generated by encrypting the User ID with a private key held by the Ministry of Health (MOH), Singapore. It can only be decrypted by MOH, and does not reveal your identity or the other person's identity. Data about devices near users is stored securely on users' device and will only be shared with MOH if users test positive for COVID-19, for the sole purpose of contact tracing. The anonymized Bluetooth data stored on your device after 25 days are automatically deleted. Therefore the right to be forgotten as per GDPR is preserved. Other third-party services will not be able to track users' identities. The Temporary ID that users' device exchanges with nearby devices are refreshed at regular intervals. The lack of a persistent identifier means it is impossible for third parties to identify or track users. Users may request for users' identification data to be deleted on their servers unless users' proximity data has already been uploaded as a confirmed case. They will then delete users' contact/mobile number, identification details and User ID from their server. This renders all meaningless data that users' device has exchanged with other devices, because that data will no longer be associated with the users.

Users' data will only be used for COVID-19 contact tracing. Any data shared with MOH will only be used solely for contact tracing of persons possibly exposed to COVID-19. TraceTogether will only communicate with nearby devices for a limited time. They will only use TraceTogether for contact tracing. Once contact tracing ceases, users will be prompted to disable the functionality of the TraceTogether App or return/dispose of the Token. For the App, users can also disable its functionality any time by turning the App's Bluetooth permissions off or deleting the App. If contact tracing is required for a future outbreak, users will be prompted to enable permissions, or users can reinstall the App. They use anonymized data to improve TraceTogether. The TraceTogether app collects anonymized data about users' phone and app, e.g., device model, app version, etc. to help it improve and provide a better user experience.

Next from Asia, particularly from the Middle-East area, Bahrain uses an application named BeAware [20]. This application needs to have good documentation mentioning the data privacy issue. Tetamman [21] in Saudi Arabia always uses GPS and Bluetooth to cause power leakage, i.e., battery drain. It takes much control over users' smartphones as well. Need to mention good documentation about data privacy. Hamagen (or "Protector" in Hebrew) [22] in Israel cross-checks the GPS history of patient's mobile phones with historical geographic data of patients from the Ministry of Health (MOH).

Meanwhile, from Europe, Czech Republic is using an application named e-Rouŝka [23] which has a proper documentation [24] with a self-explanatory video as well. But those are in Czech language. So it has a linguistic barrier. These kinds of information should be stored using international languages, such as English. In Hungary, VírusRadar [25] is a mobile app which implemented for Apple iOS and also for Android. It provides the highest security standards, full control over our personal data, and also guarantees privacy protection. It is generally using the Bluetooth Low Energy (BLE) protocol to detect faceless encrypted contacts [26] which are highly secured.

After that, from Europe, more specifically from Northern Europe, Rakning C-19 app [27] in Iceland collects the GPS location of the users' phone and stores information locally on the device. If the phone owner is diagnosed with the Covid-19 disease, then he is asked by the Health Directorate to share the location data for contact tracing to mark out the individuals that might need to go into quarantine. Meanwhile, Smittestopp [28] in Northern Europe, more precisely, Denmark, always turns GPS and Bluetooth on and drains the device battery. Frequently asked questions (FAQ) section is helpful, but not enough proper documentation was found regarding the data usage policy. Linguistic barrier is also an issue.

After that, from South America, Colombia uses CoronApp [29]. The documentation of this application will be found in [30]. Nothing has been mentioned about data privacy and further disposal of data, so it needs to mention the data privacy of an individual. Also, it needs to make an English version of documentation and existing video.

Finally, from Oceania, Australia is using an application named COVIDsafe [31]. It already has documentation [32]. Another country from this continent, New Zealand uses NZ Covid Tracer App [33] for preventing and controlling their corona situation. This is based on user interaction. A user needs to share his/her credential with the application if he/she is found to have COVID-19. This is pretty much based on users. If they don't want to share credentials, nothing can be done. But it has better data privacy though it's not as effective as other contact tracing applications. It has two-factor authentication. Data is encrypted before sent. Finding this app is difficult, and it has poor design.

## IV. SUMMARY OF CONCERNS

The following paragraphs have summarized the trust concerns detected and categorized by our through analytical procedure.

However, from those numerous anonymous datasets, a person can be identified using a demographic variable, which create a big concern about the privacy and ethical use of those data. Recently the collection and subsequent sale of Facebook data to Cambridge Analytica has made an enormous trust issue.

In 2016, journalists re-identified politicians in an anonymized browsing history dataset of three million German citizens, uncovering their medical information and their sexual preferences. A few months before, the Australian Department of Health publicly released de-identified medical records for 10% of the population only for researchers to re-identify them six weeks later. Before that, studies had shown that de-identified hospital discharge data could be re-identified using basic demographic attributes and that diagnostic codes, year of birth, gender, and ethnicity could uniquely identify patients in genomic studies data. Finally, researchers were able to uniquely identify individuals in anonymized taxi trajectories in NYC, bike sharing trips in London, subway data in Riga, and mobile phone and credit card datasets [34][35].

There has some model on re-identification of a person from an anonymous dataset. Quasi-identifier is one of it.it consists one or more quasi-identifier which called qids. Using this model a person can be identified from an anonymous dataset.

## V. RECOMMENDATIONS

Applications which are being used in different countries are not creating documentations properly. If an application has proper documentation, then people will feel more comfortable to use that application because they already know how the application in exactly working.

To increase its trust issues they should make a proper documentation and a self-explanatory video to let the people know about the working procedure of the application.

Moreover, linguistic barrier is a major concern in European countries. Many foreign personals living in European countries may not speak or read the native language, so having app tutorial in the native language causes inconvenience. This also applies for foreign volunteers working in developing countries.

Aware mass level people regarding security model such that they know that not all actors are malicious and there are very few nefarious actors. Also, a pure honest model is not possible, so *caveat emptor* (Latin, meaning "let the buyer beware") is the best policy in a semi-honest model. Moreover, even though the re-identification is mathematically possible, proper data management can reduce the threat significantly.

Furthermore, as every contact tracing application is using a centralized database to store data, if the distributed ledger is used, it will be safer for the data stored. If done so, in case of malicious attacks, the actor will be able to retrieve maybe a part of the database rather than the whole database.

Therefore, Blockchain technology can play a vital role in this regard. It ensures maximum security with a shared ledger system, which makes it the best match to address the issues raised by the centralized systems.

## VI. CONCLUSION

Researchers have identified different psychological factors associated with human behavior during this COVID-19 pandemic and the previous disease outbreak, such as panic buying and patient history hiding [36][37]. These control the individual mental health of the entire community, society and professionals, such as health care workers and persons working in clinics [38][39]. Therefore, it is highly essential to study an individual's trust issues using digital health monitoring technologies, such as health apps.

If the near future does not solve the trust issues, it will cost more as people will be reluctant to use health apps and make a pandemic to come a much worse one. Further studies and implementations have to be carried out as per the suggestions presented in the recommendations section. Only then we can ensure a better strategy to fight back future zoonosis or such.